\documentclass{iaus}
\usepackage{graphicx}

\title[VLBI Tests of GR] 
{Recent VLBA/VERA/IVS Tests of \\ General Relativity}

\author [Fomalont et al] {Ed Fomalont$^1$, Sergei Kopeikin$^2$, Dayton Jones$^3$,\\ Mareki Honma$^4$ \& Oleg Titov$^5$}   

\affiliation{$^1$ National Radio Astronomy Observatory, \\ 520 Edgemont Rd,
Charlottesville, VA 22903, USA \\ email: {\tt efomalon@nrao.edu}
\\[\affilskip]
$^2$ Dept. of Physics \&Astronomy, University of Missouri-Columbia, \\ 
223 Physics Bldg., Columbia, MO 65211, USA \\email: {\tt kopeikins@missouri.edu}
\\[\affilskip]
$^3$ Jet Propulsion Laboratory,\\ 
4800 Oak Grove Ave, Pasadena, CA 91109, USA \\email: {\tt dayton.jones@jpl.nasa.gov}
\\[\affilskip]
$^4$ VERA Project Office, Mizusawa VLBI Observatory,\\ 
NAOJ, 181-8588, Tokyo, Japan \\email: {\tt mareki.honma@nao.ac.jp}
\\[\affilskip]
$^5$ Geoscience, Australia, GPO Box 378, Canberra, ACT 2601, Australia\\ email: {\tt Oleg.Titov@ga.gov.au}}

\pubyear{2009}
\volume{xxx}  
\pagerange{119--126}
\jname{Title of your IAU Symposium}
\editors{A.C. Editor, B.D. Editor \& C.E. Editor, eds.}
\begin{document}

\maketitle

\begin{abstract}

We report on recent VLBA/VERA/IVS observational tests of General
Relativity.  First, we will summarize the results from the 2005 VLBA
experiment that determined gamma with an accuracy of 0.0003 by
measuring the deflection of four compact radio sources by the solar
gravitational field.  We discuss the limits of precision that can be
obtained with VLBA experiments in the future.  We describe recent
experiments using the three global arrays to measure the aberration of
gravity when Jupiter and Saturn passed within a few arcmin of bright
radio sources.  These reductions are still in progress, but the
anticipated positional accuracy of the VLBA experiment may be about
0.01 mas.


\end{abstract}

\firstsection 
\section{The VLBA 2005 Solar-Bending Experiment}

The history of the experiments to measured the deflection of light by
the solar gravitational field is well-known.  The 1919 optical
experiment during a solar eclipse in Brazil established the viability
of GR, and \cite[Shapiro (1964)]{sha64} suggested that radio
techniques would provide much better accuracy.  The precision of radio
experiments over the last 40 years $\gamma< 0.001$ was achieved
(e.g.~\cite[Lebach \etal\ 1995]{leb95}, \cite[Shapiro \etal\
2004]{sha04}).

\begin{figure}[t]
\begin{center}
 \includegraphics[width=4.5in]{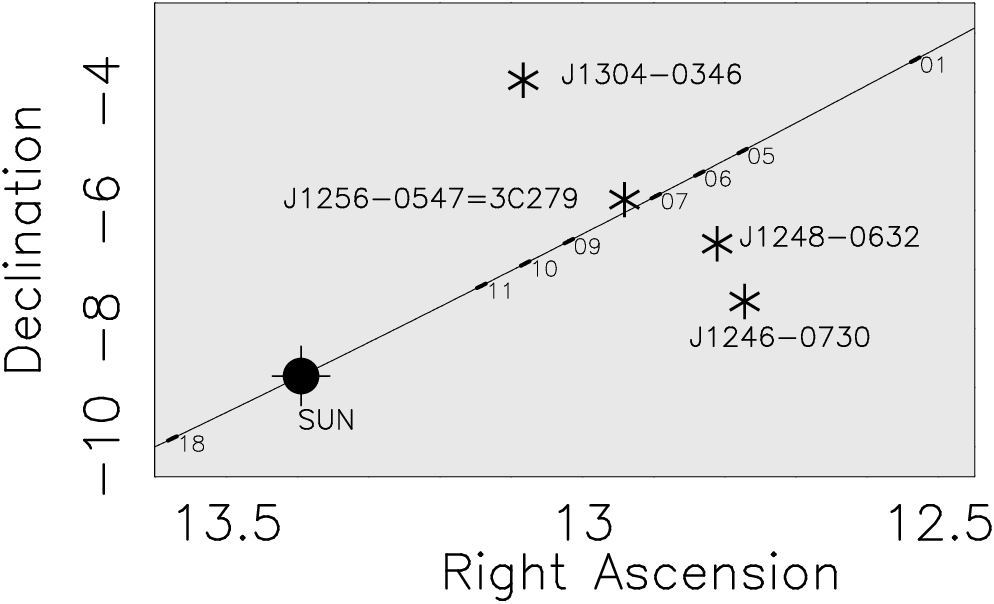} 

\caption{{\bf The Source Configuration for the Solar Deflection
Experiment of 2005:} The solar trajectory between October 1 and 18 is
shown by the diagonal line, with the eight observing days
superimposed.  The location of the four radio sources are indicated.}
\label{fig1}
\end{center}
\end{figure}

Since the inception of operation in 1990, the positional accuracy that the
Very Long Baseline Array (VLBA) can achieve has increased markedly,
and it is now capable of measuring the angular separation between
compact radio sources that are a few degrees apart in the sky with an
accuracy of about 0.01 mas.  These advances were made from
increased stability and sensitivity of the array, more accurate
determination of astrometric and geodetic parameters from the
International VLBI Service (IVS), and the use of phase delay
referencing even with antenna baselines greater than 5000 km.  We,
thus, observed on eight days in October 2005 in order to determined
the change in relative position among four sources as the sun moved
through the area in order to measure $\gamma$ with high precision.
The observing days and sources are shown in Fig.~ \ref{fig1}.

In order to minimize the effects of the solar corona, we used the
highest routinely used VLBA frequency of 43 GHz.  In order to lessen
the effects of the changing tropospheric refraction above each
telescope, we alternated source observations every 40 second.  The
choice of sources was a compromise among three criteria: sufficiently
strong and compact to be detectable by the VLBA in about 20 sec;
sufficiently close in the sky to avoid large tropospheric phase
changes; but not closer than about two degrees in order to have a
significant differential solar gravitational bending.  We also
observed at 23 GHz and 15 GHz (switching frequencies every 40 min) in
order to estimate and remove the long-term coronal refraction.

The description of the experimental parameters, the reduction methods,
and the analysis technique to estimate $\gamma$ and its uncertainty
are given in \cite[Fomalont \etal\ (2009)]{fom09}.  The results are
$\gamma = 0.9998\pm 0.0003$ (standard error).  This is the most
accurate radio interferometric result to date, although it is less
accurate than the 2002 Doppler-tracking Cassini experiment
(\cite[Bertotti \etal\ 2003]{ber03}) if one postulates that the
translational gravito-magnetic field affects propagation of radio
waves exactly as predicted in general relativity.  This postulate was
not parameterized in NASA ODP code and could not be directly tested in
the Cassini experiment (\cite[Kopeikin \etal\ 2004]{kop07},
\cite[Bertotti \etal\ 2008]{ber08}, \cite[Kopeikin \etal\
2009]{kop09}).

The accuracy of future VLBA experiments can be increased by a factor of
four with several improvements: observing on all days when the sources
were between $4^\circ$ to $7^\circ$ from the sun; choosing available
sources in April to August when the sun is at its most northern
declinations; increasing the relative observing time at 43 GHz since
the coronal refraction correction was small; and scheduling an
experiment as often as possible since about ten groups of sources near
the ecliptic are available for a high precision experiment.

\section {The Planetary Gravitational Aberration Experiments}

On September 8, 2002, Jupiter passed within $4'$ of the quasar
J0842+1835.  Because of the motion of Jupiter, the gravitational
bending of the quasar position was not precisely radial from the
planet. For this experiment, the GR prediction of the non-radial
deflection is 0.05 mas in the direction of Jupiter's motion in the sky
at closest Jupiter/quasar encounter. (The radial deflection component
at this time was 1.1 mas.)  This aberrational-type deflection varies
as $(v_J/c)~d^{-2}$, where $d$ is the angular separation in the sky
between Jupiter and the quasar, and $v_J$ is the {\it heliocentric}
velocity of Jupiter (\cite[Kopeikin 2001]{kop01}).  This aberrational
deflection was measured with the VLBA with an accuracy of 20\%
(\cite[Fomalont \& Kopeikin 2003]{fom03}). It demonstrated the
gravito-magnetic effect caused by translational mass-current of a
moving body, and also initiated many discussions about the the
experimental interpretation in the framework of the general theory of
relativity.

\begin{figure}[t]
\begin{center}
 \includegraphics[width=4.5in]{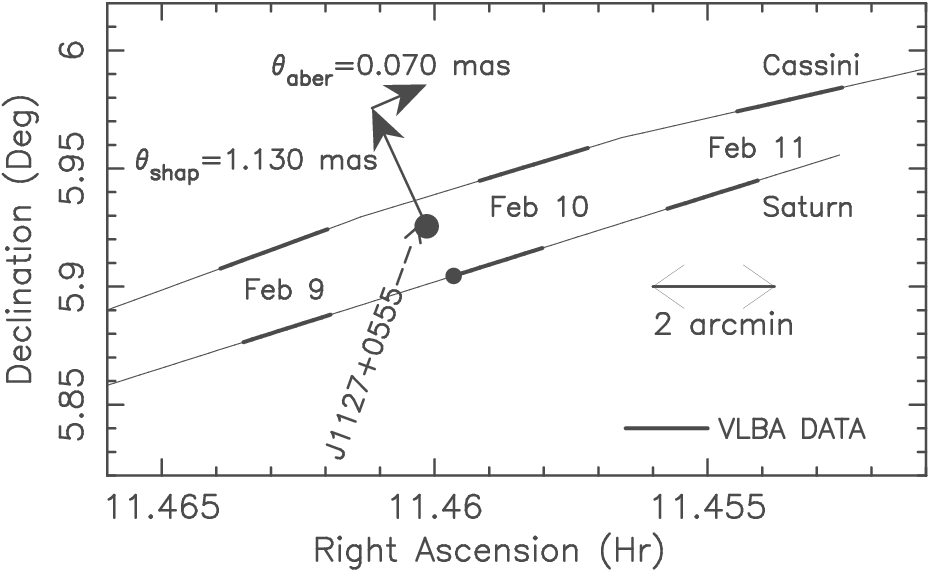} 
\caption{{\bf The Source Configuration for the February 2009
Aberration Experiment:} The sky motion of Saturn and the Cassini
spacecraft are shown by the lines from the lower left to upper right.
The VLBA observing periods are shown by the bold part of the lines.
At the closest approach of Saturn to J$1127+0555$, the
predicted source deflection is 1.13 mas and the aberrational
deflection is 0.070 mas.  }

\label{fig2} 
\end{center} 
\end{figure}

Two planetary near encounters with bright radio sources have recently
occurred.  On November 19, 2008, Jupiter passed within $1.4'$ of
J$1925-2219$, and on February 10, 2009, Saturn pass within $1.3'$ of
J$1127+0555$.  The arrays that observed these encounters were:
International VLBI Service (IVS) array for both dates, the Japanese
VLBI Exploration of Radio Astronomy (VERA) array for the November
encounter (\cite[Honma \etal\ 2003]{hon03}), and the VLBA for the
February encounter.  The IVS observing programs are coordinated with
the international community and use a variety of telescopes for
semi-weekly to bi-monthly observations to monitor the earth
orientation (see http://www.iers.org).  For the November 19
experiment, the observing array included Parkes, Hobart26, Kokee, and
Tsukub32. The nominal IVS 24-hour schedule (session OHIG60) was
modified so that additional 8-hour observations of J$1925-2219$ and a
nearby source J$2000-1748$ were included in order to determine the
change of their relative position during the experiment.  For the
February 10 experiment (session RD0902), the observing array consisted
of Parkes, Hobart26, Medicina, Matera, Badary, LA-VLBA, and
KP-VLBA. The calibrator source for J$1127+0555$ was J$1112+0724$.
The analysis of these two experiments are progressing.

The VERA observations were made on November 17, 19 (day of close
encounter) and 22, each day for seven hours.  This array has a
dual-beam system so that two sources within $2.2^\circ$ can be
observed simultaneously---J$1923-2104$ and J$1925-2219$ are separated
by $1.4^\circ$---and the expectation is that their relative positions
can be determined to an accuracy of about 0.05 mas every two hours.
The analysis is also in progress.

\begin{figure}[t]
\begin{center}
 \includegraphics[width=4.2in]{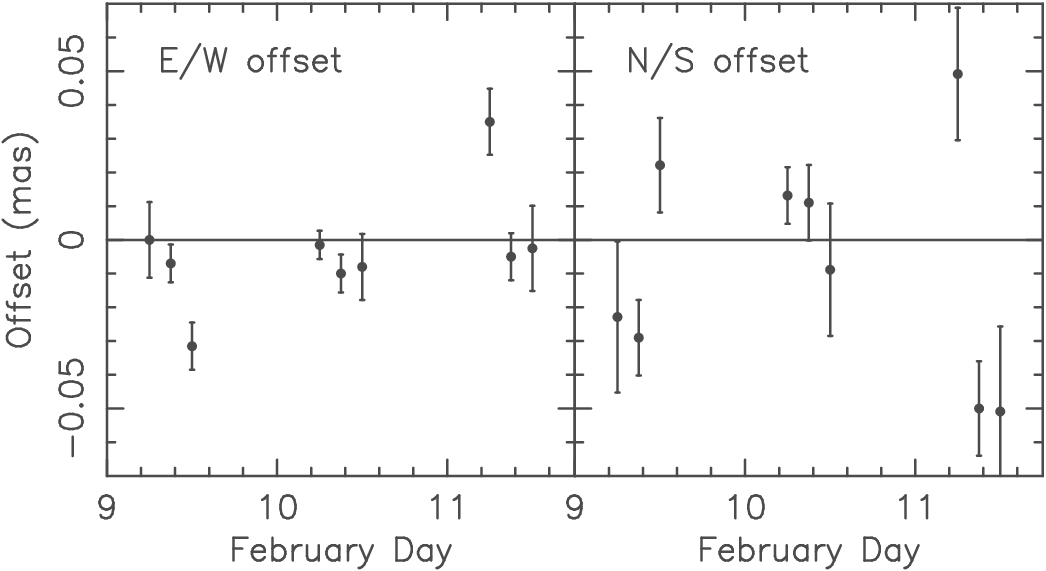} 

\caption{{\bf The Residual Differential Separation Between Cassini and
J1127+0555:} The separation between the two sources, after removal
of the best-fit offset, velocity and acceleration from the measured
offsets.  The predicted gravitational bending of both Cassini and
J$1127+0555$ has been removed, so the difference should be zero if the
GR prediction is correct.  The left plot is the east-west difference,
the right plot is the north-south difference.  The position difference
estimates were obtained for the three 2.5-hour intervals on each day,
and the error are the one-sigma errors expected from signal-to-noise
considerations alone.
 }

\label{fig3} 
\end{center} 
\end{figure}

The reductions for the February VLBA experiment are nearly completed.
The source configuration during the experiment is shown in Fig.~\ref
{fig2}.  An interesting aspect of this experiment is that
J$1127+0555$ was measured with respect to Cassini. (The emission
from Saturn is too extended to be detected by the VLBA.) which was
within $5'$ of the source between February 9 to 11.  With such a close
encounter, the effects of the tropospheric refraction are small so the
relative position of Cassini and the source could be measured with
high accuracy, with the limit imposed by the signal-to-noise of the
experiment of $0.004$ mas.  Both objects were sufficiently close to be
observed simultaneously, rather than by switching between sources,
further decreasing the tropospheric effects.  However, the measurement
of the deflection of the source requires that the orbit of Cassini
is precisely known.

Because the VLBA experiment was correlated in early March using an
approximate orbit of Cassini, the assumed position of Cassini used in
the reductions produced a position difference with that of J$1127+0555$
that slowly drifted during the 3-day experiment.  Cassini's orbital
parameters cannot be predicted more than one month in advance because
of interactions with the Saturnian moons and also by the occasional Cassini
thrusts to optimize the orbit.  We expect to have a much more accurate orbit
for Cassini from the JPL Cassini navigation group by mid-2009.

Nevertheless, it is possible to surmise the positional sensitivity of
the VLBA experiment.  A good assumption is that the orbital model
error of Cassini used in the VLBA reductions can be approximated by
constant offset, velocity and acceleration over the observing period.
If we further remove the radial and aberration deflection prediction
by GR, then the resultant relative position of Cassini with respect to
J$1127+0555$ should be zero.  This position difference with the above
adjustments is shown in Fig.~\ref{fig3}.  The departure of the
residuals from zero has an error of about 0.01 mas E/W, and 0.02 mas
N/S (this resolution is twice as poor).  Since the aberrational
deflection is 0.07 mas at closest encounter on February 10, this
experiment may be a more accurate measure of the aberration deflection
than that of the 2002 Jupiter experiment \cite[Fomalont \& Kopeikin
2003]{fom03}.

For the actual position comparison, we will use the observations on
February 9 and 11 to determine the residual offset and velocity of
Cassini with respect to J$1127+0555$.  The acceleration residial of
the Cassini orbit, however, must be known to less than $2.6\times
10^{-6}$~m$^{-2}$, corresponding to angular change of 0.02 mas over 48
hours, in order to interpolate the spacecraft quasar offset accurately
on February 10 when the gravitational deflection is large.

\section {Conclusion}

Using the VLBA at 43 GHz with phase referencing observation, we have
measured $\gamma$ with an rms precision of 0.0003.  With improved
experimental strategies and observations of many source clusters near
the ecliptic, the precision can be increased by at least a factor of
four.

Recent experiments with the VLBA, the IVS array and VERA of radio
sources with near encounters with Jupiter and Saturn will provide more
accurate measurements of the gravito-magnetic effect by measuring the
aberrational deflection of the sources.  Results are not yet
available, although the VLBA precision may be about 0.01 mas and,
thus, produce a more accurate result than the 2002 VLBA experiment
(\cite [Fomalont \& Kopeikin 2003]{fom03}).

The National Radio Astronomy Observatory is a facility of the National
Science Foundation operated under cooperative agreement by Associated
Universities, Inc.  Part of this research was performed at the Jet
Propulsion Laboratory, California Institute of Technology, under
contract with the National Aeronautics and Space Administration.  The
Parkes 64-meter is operated by the Australian Telescope National
Facility (ATNF). The scheduling of the both IVS sessions was kindly
done by Dirk Behrend and John Gipson (NVI GSFC). Sergei Kopeikin
thanks the Research Council (Grant No. C1669103) and Alumni
Organization of the University of Missouri-Columbia (2009 Faculty
Incentive Grant) for support of this work.

\end{document}